\documentclass{arxiv}
\usepackage{cite}
\usepackage{txfonts}
\usepackage{graphicx}% 画像の読み込み用のパッケージ
\usepackage{subfigure}%
\newcommand{\bi}[1]{
\ensuremath{\boldsymbol{#1}}}

\title{Ensemble Inequivalence in the Spherical Spin Glass Model with  Nonlinear Interactions}

\author{Yuma Murata and Hidetoshi Nishimori  %\\
}
\inst{\address{Department of Physics, Tokyo Institute of Technology, Tokyo 152-8551, Japan} %\\
}
\abst{ We investigate the ensemble inequivalence of the spherical spin glass model with nonlinear interactions of polynomial order $p$. This model is solved exactly for arbitrary $p$ and is shown to have first-order phase transitions between the paramagnetic and spin glass or ferromagnetic phases for $p \geq 5$. In the parameter region around the first-order transitions, the solutions give different results depending on the ensemble used for the analysis. In particular, we observe that the microcanonical specific heat can be negative and the phase may not be uniquely determined by the temperature.}
\kword{ensemble inequivalence, long-range interacting systems, classical spin systems, disordered systems, spin glasses, phase diagrams}

\begin{document}
\pagestyle{empty}
\maketitle\thispagestyle{empty}

\section{Introduction}
Statistical ensembles play key roles in the investigation of systems with a large number of particles. The most fundamental one is the microcanonical ensemble, and it is believed that we can derive other ensembles from the microcanonical ensemble.\cite{1} Alternatively, we can relate the microcanonical ensemble to the canonical ensemble by the Legendre transformation.\cite{1} It implies that the physical properties of a system do not depend on the statistical ensemble in the thermodynamic limit. In other words, these ensembles are generally considered to be equivalent. This fact enables us to choose a statistical ensemble which is useful for practical calculations, and the canonical ensemble is usually the first choice. However, this property is known to hold only for short-range interacting systems. For long-range interacting systems, we have to reconsider whether or not ensembles are equivalent.

In fact, the phenomena that the physical properties depend on statistical ensemble emerge in systems with long-range interactions. In particular, for the systems with first-order transitions, ensemble in-equivalence has been reported in previous studies.\cite{2,3,4,5,6,7,8,9,10,11,12,13,14,15,16} An interaction decaying in a power low $r^{\alpha}~(\alpha \le d)$, where $r$ is the distance between particles and $d$ is the spatial dimension, is said to be long ranged.\cite{5,6,7,8} Long-range interacting systems with first-order phase transitions have a striking feature that the specific heat can be negative under the micro-canonical ensemble. Such ensemble inequivalence has traditionally been discussed mainly in astrophysics.\cite{9,10,11,12}

Recently, these problems outside of astrophysics have been analyzed from the view point of ensemble inequivalence, results of which have further promoted the understanding of ensemble inequivalence. Many non-trivial results related to ensemble inequivalence have been reported in spin systems without disorder$^{13-17)}$ and driven systems with local dynamics.$^{18,\ 19)}$ Analyses of long-range interacting spin systems with disorder have only recently been initiated,$^{20-23)}$ but our knowledge on ensemble inequivalence in systems with spin glass transitions is still limited.\cite{20,22,23} Therefore,  it is an interesting problem whether or not ensemble inequivalence are observed in the other spin systems with disorder.

In this paper, we analyze the spherical spin glass model with non-linear interactions both by the canonical and microcanonical ensembles, to investigate whether or not ensemble inequivalence emerge in this spin glass system. This model is a generalization of the linear model introduced by Kosterlitz et al$^{24,\ 25)}$ in conjunction with the nonlinear interaction for the non-disordered cases$^{17,\ 26-29)}$ and has the advantage that all calculations can be done exactly. We can thus avoid the replica method and do not have to worry about the limit of applicability of the replica method. 

The organization of this paper is as follows. The model is defined in $\S$ 2. In $\S$ 3, we calculate the partition function in the canonical ensemble (the number of states in the microcanonical ensemble) and obtain the free energy density (the entropy density). We analyze the behavior of thermodynamic functions and phase diagrams in $\S$ 4. The last section is devoted to summary and conclusion.

\section{The Model}
Let us introduce the spherical model of spin glasses with non-linear interactions. It is defined by the Hamiltonian
\begin{eqnarray}
H=-NV\left( \frac{1}{N}\sum^{N}_{i<j} \xi_{ij}S_{i}S_{j}\right). 
\end{eqnarray} 
Here, $V\left(x\right)$ is given by
\begin{eqnarray}
V\left( x \right) = \frac{1}{2^{p-1}p}\left[ \left( 1+x \right)^{p} -1 \right],
\label{eq.v}
\end{eqnarray}
and $\xi_{ij}$ is a dimensionless random parameter, whose probability distribution $P\left(\xi_{ij}\right)$ is Gaussian with average $\xi_{0}/N$ and variance $1/N$,
\begin{eqnarray}
P \left( \xi_{ij} \right) = \sqrt[]{ \frac{N}{2\pi}\mathstrut} \exp \left[ -\frac{N}{2}\left( \xi_{ij}-\frac{\xi_{0}}{N}\right)^{2} \right].
\end{eqnarray}
The spin variables $S_{i}$ obey the spherical constraint
\begin{eqnarray}
{\bi S}^{2} = \sum_{i=1}^{N}S_{i}^{2} = N.
\end{eqnarray}

In the case of $p=1$, $V\left(x\right)$ is equal to $x$ and the model reduces to the spherical model introduced by Kosterlitz et al.\cite{24} 

In the absence of disorder\cite{26,27,28,29}, the system shows first-order transitions and ensemble equivalence for $p \geq 5$ as long as the interactions are short-ranged.\cite{17}
\section{Free energy and entropy}
In this section, we calculate the partition function and the number of states, from which we derive the canonical free energy and the microcanonical entropy. 
\subsection{Free energy in the canonical ensemble}
First we solve the model in the canonical ensemble. The partition function of the system can be written as
\begin{eqnarray}
Z &=& {\rm Tr} \exp\left[ N \beta V\left( \frac{1}{N}\sum^{N}_{i<j} \xi_{ij}S_{i}S_{j}\right) \right] \delta \left( {\bi S}^{2}-N \right),
\end{eqnarray}
where $\beta$ is the inverse temperature and the trace denotes integrations over the spin variables. We introduce the auxiliary variables $z$, $\rho$ and $\lambda$ to facilitate the integration, 
\begin{eqnarray}
Z &=&\int_{}^{}dz d\rho d\lambda\ {\rm Tr}\ \exp\left[ N \beta V\left( \rho \right) -z\left(  \sum^{N}_{i=1}S_{i}^{2}-N\right) - \lambda \left( N \rho-\sum_{i<j}\xi_{ij}S_{i}S_{j} \right)\right] \nonumber \\
&=& \int_{}^{}dz d\rho d\lambda \exp \left[ N\beta V\left( \rho \right) +N z - N\lambda  \rho + \log {\rm Tr}\ \exp\left(  -z \sum^{N}_{i=1}S_{i}^{2} + \lambda  \sum^{}_{i<j} \xi_{ij}S_{i}S_{j}\right) \right]. \nonumber \\  
\end{eqnarray}
By diagonalizing the random matrix $\xi_{ij}$, we calculate the trace part as a Gaussian integral. If we denote the eigenvalues of the random matrix $\left\{ \xi_{ij} \right\}$ as $\left\{ \xi_{\alpha} \right\}$, we can write
\begin{eqnarray}
Z &=& \int dz d\rho d\lambda \exp \left\{ N \left[  \beta V\left( \rho \right) +z - \lambda \rho- \frac{1}{2N}\sum_{\alpha}\log \left( z-\frac{1}{2}\lambda \xi_{\alpha}\right) \right] \right\}.
\end{eqnarray}
The saddle point equations for the evaluation of the integral for large $N$ are
\begin{eqnarray}
1 &=& \frac{1}{N}\sum_{\alpha}\frac{1}{2z-\lambda \xi_{\alpha}}, \\
\lambda &=& \beta \frac{\partial V}{\partial \rho}, \\
\rho &=& \frac{1}{2N}\sum_{\alpha}\frac{\xi_{\alpha}}{2z-\lambda \xi_{\alpha}}.
\end{eqnarray}
In the thermodynamic limit, the summation over $\alpha$ is replaced by an integral,
\begin{eqnarray}
\frac{1}{N} \sum^{}_{\alpha}  \rightarrow  \int_{-\infty}^{\infty}d\xi_{\alpha} \mu\left( \xi_{\alpha}\right). 
\end{eqnarray}
The density of eigenvalues $\mu(\xi_{\alpha})$ can be expressed as 
\begin{eqnarray}
\mu \left( \xi_{\alpha} \right) = \left\{ \begin{array}{ll}
\mu_{0} \left( \xi_{\alpha} \right) \ \ \left( \xi_{0} \leq 1\right) \\
\mu_{0} \left( \xi_{\alpha} \right) + \frac{1}{N}\delta\left( \xi_{\alpha} - \xi_{m} \right) \ \ \left( \xi_{0} > 1\right) . \\
\end{array} \right.
\end{eqnarray}
Here $\mu_{0}\left(\xi_{\alpha}\right)$ obeys the semicircular law $\sqrt[]{\mathstrut 4-\xi_{\alpha}^{2}}/2\pi$.$^{30)}$ If $\xi_{0} > 1$, the eigenvalue spectrum is modified by an isolated eigenvalue $\xi_{m}=\xi_{0}+1/\xi_{0}$.$^{31)}$ The thermodynamic function for $\xi_{0}>1$ describes the ferromagnetic (FM) phase. 

Using eqs. (11) and (12), we evaluate the saddle point equations, eqs. (8) and (10), to determine the state of the system. We eliminate the auxiliary variables $\lambda$ and $\rho$ from the saddle point equations. Then the free energy density $f= -T\log Z/N$ and the inverse temperature $\beta$ of each phase are given as functions of $z$, from which $f$ is determined as a function of $\beta$. Note that we should carefully treat the estimation of saddle point equations when we replace the summation of eigenvalues by an integral. Details are described in Appendix.  

The paramagnetic (PM) and spin glass (SG) phases exist when the ferromagnetic bias is small, $\xi_{0} \leq 1$. The free energy density and the inverse temperature of the PM phase are   
\begin{eqnarray}
-\beta_{\rm PM} f_{\rm PM} &=& \beta_{\rm PM} V\left( \frac{\sqrt[]{2z-1\mathstrut}}{2} \right) - \frac{1}{2}z+\frac{1}{2}\log2 + \frac{3}{4},  \\
\beta_{\rm PM}  &=& \frac{2^{p-1}\sqrt[]{2z-1\mathstrut}}{\left( 1+\sqrt[]{2z-1\mathstrut}/2\right)^{p-1}},
\end{eqnarray}
which are valid for $1/2 \leq z \leq 1$. It is seen in eq. (14) that $z=1/2$ corresponds to $\beta = 0\ \left( T \rightarrow \infty \right)$, which justifies us to identify this range of $z$ with the PM phase. The SG phase exists when $z>1$. The free energy and the inverse temperature are given by
\begin{eqnarray}
-\beta_{\rm SG} f_{\rm SG} &=& \beta_{\rm SG} V\left( 1- \frac{1}{2z} \right) -\frac{1}{2}\log z+\frac{1}{2}\log2 +\frac{1}{4},  \\
\beta_{\rm SG} &=& \frac{2^{p-1}z}{\left(2-1/2z \right)^{p-1}}.
\end{eqnarray}
We can confirm that the free energy densities of the PM and SG phases take the same value at $z=1$, which agrees with the previous result for $p=1$.$^{24,\ 25)}$ 

The free energy density and the inverse temperature of the FM phase for $\xi_{0} >1$ are given by   
\begin{eqnarray}
-\beta_{\rm FM} f_{\rm FM} &=& \beta_{\rm FM}  V\left[\frac{\xi_{m}}{2}\left( 1- \frac{1}{2z} \right)\right] -\frac{1}{2}\log z+\frac{1}{2}\log2 +\frac{1}{4} + h\left(\xi_{m}\right), \\
\beta_{\rm FM} &=& \frac{2^{p}z}{\xi_{m} \left[1+\xi_{m}\left( 1- 1/2z \right) /2 \right]^{p-1}}, 
\end{eqnarray}
where
\begin{eqnarray}
h\left(\xi_{m}\right) &=&  -\frac{\xi_{m} \left(  \xi_{m}-\sqrt[]{\mathstrut \xi_{m}^{2}-4}\right)}{8}-\frac{1}{2}\log \frac{\xi_{m}+\sqrt[]{\mathstrut \xi_{m}^{2}-4}}{\xi_{m}}+\frac{1}{2}.
\end{eqnarray}
Notice that the free energy density and the inverse temperature of the FM phase are equal to those of the SG phase when $\xi_{0} =1$ corresponding to $\xi_{m}=2$. From this fact, we confirm that $f\left(T \right)$ is a smooth, continuous function at transition point between the FM and SG phases for arbitrary $p$.
 %It implies that the entropy of the SG and FM phases are smooth continuous function for $\xi_{0} =1$. Therefore, the phase transition is of second-order between the SG and FM phases. %

\subsection{Entropy in the microcanonical ensemble}
We next calculate the number of states for the microcanonical ensemble as described in Appendix,
\begin{eqnarray}
\Omega &=& {\rm Tr} \delta \left(E-H\right) \nonumber \\ 
       &=&  \frac{1}{2\pi} \int dt dz d\rho d\lambda \exp \left\{ N \left[ it\left( \epsilon + V\left( \rho \right) \right)  +z - \lambda \rho - \frac{1}{2N} \sum_{\alpha} \log \left( z-\frac{1}{2} \lambda \xi_{\alpha} \right) \right]\right\},
\end{eqnarray}
where $\epsilon$ is the energy density $E/N$. The saddle point equations of the microcanonical ensemble have very similar expressions to the canonical case in eqs. (8)-(10),
\begin{eqnarray}
1 &=& \frac{1}{N}\sum_{\alpha}\frac{1}{2z-\lambda \xi_{\alpha}}, \\
\lambda &=& it  \frac{\partial V}{\partial \rho}, \\
\rho &=& \frac{1}{2N}\sum_{\alpha}\frac{\xi_{\alpha}}{2z-\lambda \xi_{\alpha}},
\end{eqnarray}
with an additional equation derived by the derivative of the exponent of eq. (20) with respect to $t$, 
\begin{eqnarray}
0 = \epsilon +  V\left( \rho \right).
\end{eqnarray}
We solve these equations and find the microcanonical entropy density $s=\log \Omega/N $ as described in Appendix. The PM and SG phases are defined for $\xi_{0}<1$. The entropy and energy densities of the PM phase for $1/2 \leq z \leq 1$ are given by  
\begin{eqnarray}
s_{\rm PM} &=& - \frac{1}{2}z+\frac{1}{2}\log2 + \frac{3}{4},  \\
\epsilon_{\rm PM} &=& -V\left( \frac{\sqrt[]{2z-1\mathstrut}}{2} \right).
\end{eqnarray}
The entropy and energy densities for $z>1$ describe the SG phase,
\begin{eqnarray}
s_{\rm SG} &=& -\frac{1}{2}\log z+\frac{1}{2}\log2 +\frac{1}{4},  \\
\epsilon_{\rm SG} &=& -V\left( 1-\frac{1}{2z}\right).
\end{eqnarray}
Similarly, the entropy and energy for the FM phase for $\xi_{0}>1$ are given by
\begin{eqnarray}
s_{\rm FM} &=& -\frac{1}{2}\log z+\frac{1}{2}\log2 +\frac{1}{4} +h\left(\xi_{m}\right),  \\
\epsilon_{\rm FM} &=& -V\left[ \frac{\xi_{m}}{2}\left(1-\frac{1}{2z}\right)\right].
\end{eqnarray} 
Here $h\left(\xi_{m}\right)$ is the same function as in eq. (19). The FM solutions are equal to the SG ones at $\xi_{0}=1$. Therefore, $s\left(\epsilon \right)$ continuously changes between the FM and SG phases.  

\section{Comparison of the thermodynamic functions and the phase diagrams}
In this section, we compare the canonical and microcanonical solutions for $p=1$ to $p=5$. We find that the phase transition between the PM and SG phases is of first order in the canonical solution for $p \geq 5$, which results in ensemble inequivalence.  
     
\subsection{Thermodynamic functions}
First, we focus our attention to the PM and SG phases ($\xi_{0} < 1$).  
\begin{figure}[htbp]
  \begin{center}
    \begin{tabular}{cc}
      \resizebox{70mm}{!}{\subfigure[The canonical inverse temperature]{\includegraphics{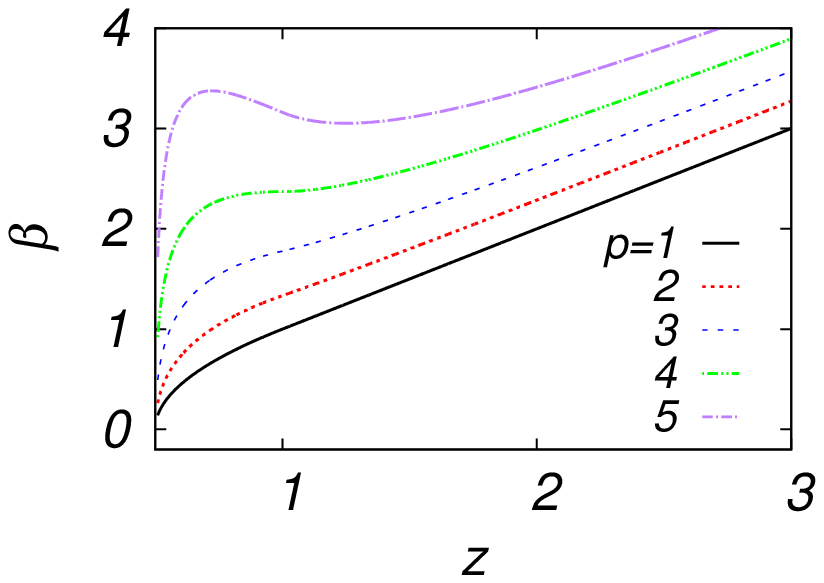}}} &
      \resizebox{70mm}{!}{\subfigure[The microcanonical energy density]{\includegraphics{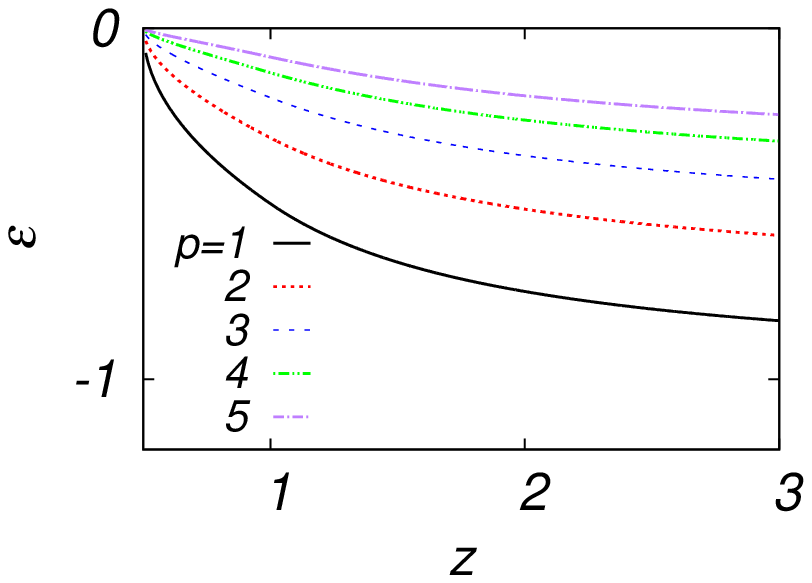}}} 
    \end{tabular}
    \caption{ The inverse temperature in the canonical ensemble in Fig. 1(a) and the energy density in the microcanonical ensemble in Fig. 1(b) for $p=1$, $2$, $3$, $4$ and $5$ as functions of $z$. When $p \leq 4$, $\beta\left( z \right)$ is a monotonically increasing function. $\epsilon\left( z \right)$ is a decreasing function for any $p$.}
    \label{figure 1}
  \end{center}
\end{figure}
In Fig. 1, we plot the inverse temperature $\beta_{\rm PM}$ and $\beta_{\rm SG}$ of eqs. (14) and (16) for the canonical ensemble in Fig. 1(a) and the energy density $\epsilon_{\rm PM}$ and $\epsilon_{\rm SG}$ of eqs. (26) and (28) for the microcanonical ensemble in Fig. 1(b) for $p=1$, $2$, $3$, $4$ and $5$. Notice that $\beta_{\rm PM}$ is for $z \leq 1$ and $\beta_{\rm SG}$ is for $z \geq 1$ in Fig. 1(a) and similarly for Fig. 1(b). In the canonical case, $\beta \left( z\right)$ is a monotonically increasing function of $z$ for $p \leq 4$. When $p = 5$, in contrast, $\beta \left( z\right)$ is not monotonic and $z$ is not uniquely determined for a given $\beta$ in a certain range. We can understand this fact as an indication that the transition between the PM and SG phases is of first order for $p = 5$. We also find that the inverse temperature of $p \geq 6$ behaves qualitatively in the same way as the one for $p=5$, which we do not show in the figure for simplicity. By contrast, $\epsilon \left( z \right)$ is a monotonically decreasing function of $z$ for any $p$ as seen in Fig. 1(b). We also observe that first-order phase transitions appear between the PM and FM phases for $p=5$ and $\xi_{0} >1$, because $\beta\left( z\right)$ is not a monotonically increasing function of $z$ (Fig. 2). The same is true for $p \geq 5$ but not $p \leq 4$. As can be understood from these results, in order to discuss the ensemble inequivalence, it is sufficient to pick up the typical cases of $p=1$ and $p=5$.  
\begin{figure}[htbp]
  \begin{center}
      \resizebox{70mm}{!}{\includegraphics{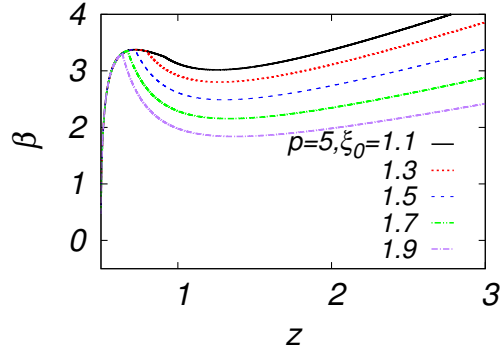}} 
    \caption{The inverse temperature for $p=5$ and $\xi_{0} = 1.1$, $1.3$, $1.5$, $1.7$ and $1.9$ as a function of $z$.}
    \label{figure 2}
  \end{center}
\end{figure}
       
Next we consider the thermodynamic functions in the PM and SG phases (i.e. $\xi_{0}<1$). We plot the free energy density $f\left( T \right)$ of eqs. (13) and (15) as a function of $T$ in Figs. 3(a) and 3(b) for $p=1$ and $p=5$.
\begin{figure}[htbp]
  \begin{center}
    \begin{tabular}{cc}
      \resizebox{70mm}{!}{\subfigure[The free energy density for $p=1$ ]{\includegraphics{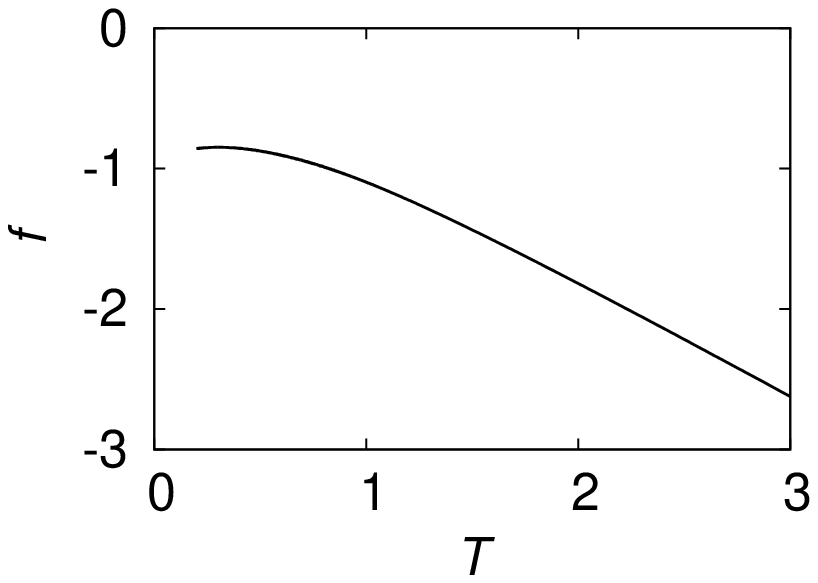}}} &
      \resizebox{70mm}{!}{\subfigure[The free energy density for $p=5$]{\includegraphics{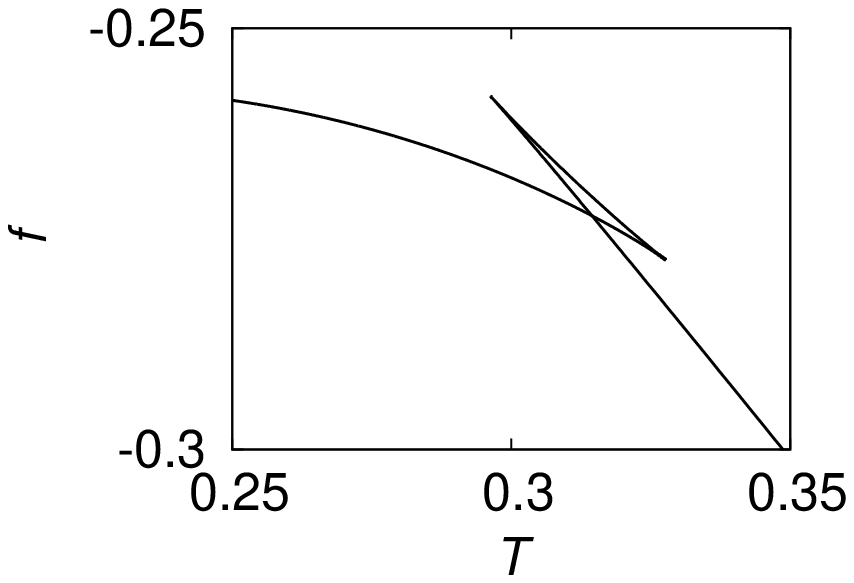}}} 
    \end{tabular}
    \caption{Free energy density $f\left( T\right)$ in the canonical ensemble for $p=1$ (Fig. 3(a)) and $p=5$ (Fig. 3(b)). }
    \label{figure 3}
  \end{center}
\end{figure}
For $p=1$, the critical temperature is $T_{{\rm c}} = 1$ as seen in eqs. (14) and (16) with $z=1$. We confirm that the phase transition is of second order because $\partial^{2} f/ \partial T^{2}$ is discontinuous at $z=1$. For $p=5$, as we can see in Fig. 3(b), the free energy density $f\left(T\right)$ shows a first-order phase transition at $T_{{\rm c}}=0.31$, which has already been expected in the previous discussion on $\beta\left(z\right)$.    
\begin{figure}[htbp]
  \begin{center}
    \begin{tabular}{cc}
      \resizebox{70mm}{!}{\subfigure[The microcanonical inverse temperature for $p=1$]{\includegraphics{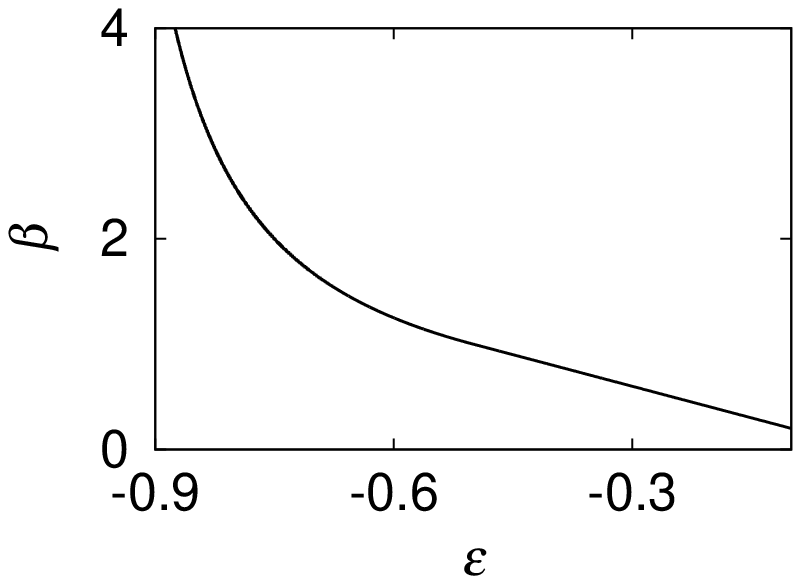}}} &
      \resizebox{70mm}{!}{\subfigure[The microcanonical entropy density for $p=1$]{\includegraphics{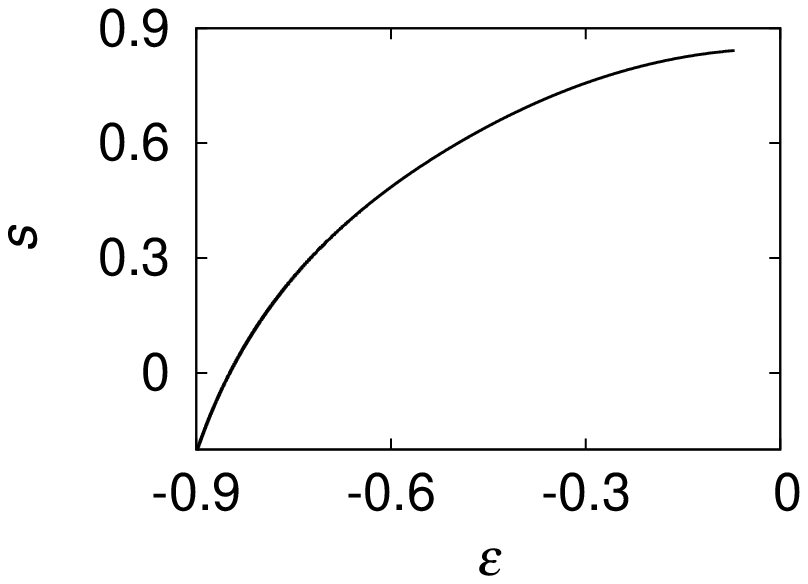}}} 
    \end{tabular}
    \caption{The inverse temperature $\beta\left(\epsilon \right)$ (Fig. 4(a)) and the entropy density $s\left(\epsilon \right)$ (Fig. 4(b)) in the microcanonical ensemble for $p=1$. A second-order phase transition exists between the PM and SG phases at $\epsilon =-0.5$.}
    \label{figure 4}
  \end{center}
\end{figure}

We also study the entropy $s\left(\epsilon \right)$ and the inverse temperature $\beta\left(\epsilon \right)$ for $p=1$ and $p=5$ under the microcanonical ensemble. As the inverse temperature $\beta\left(\epsilon \right)$ for $p=1$ is a monotonically decreasing function in Fig. 4(a), the shape of the entropy density $s\left(\epsilon \right)$ is concave (Fig. 4(b)). The phase transition between the PM and SG phases is of second order because $\partial s/\partial T$ shows a singularity at $T=1$. In this case, ensemble inequivalence does not appear and we are able to obtain the free energy, which is the same one as the canonical case, by a Legendre transformation of the microcanonical entropy.              
\begin{figure}[htbp]
  \begin{center}
    \begin{tabular}{cc}
      \resizebox{70mm}{!}{\subfigure[The microcanonical inverse temperature for $p=5$]{\includegraphics{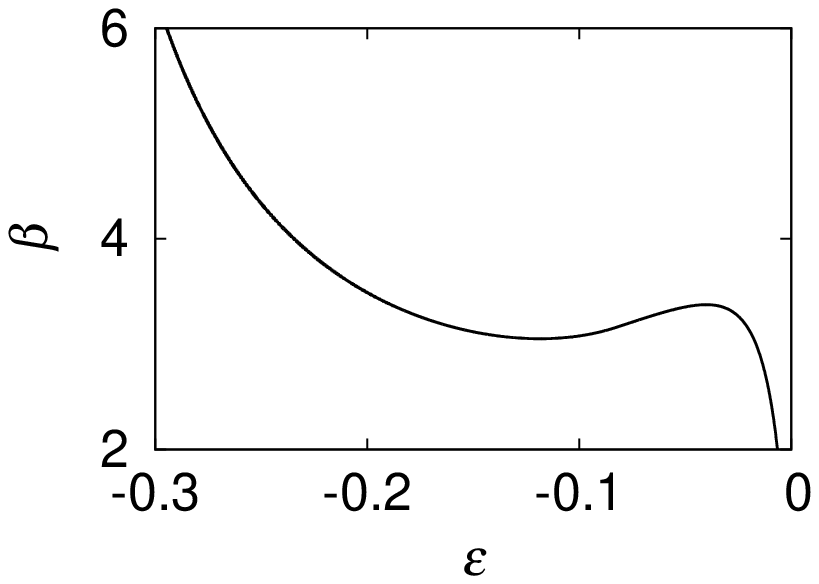}}} &
      \resizebox{70mm}{!}{\subfigure[The microcanonical entropy density for $p=5$]{\includegraphics{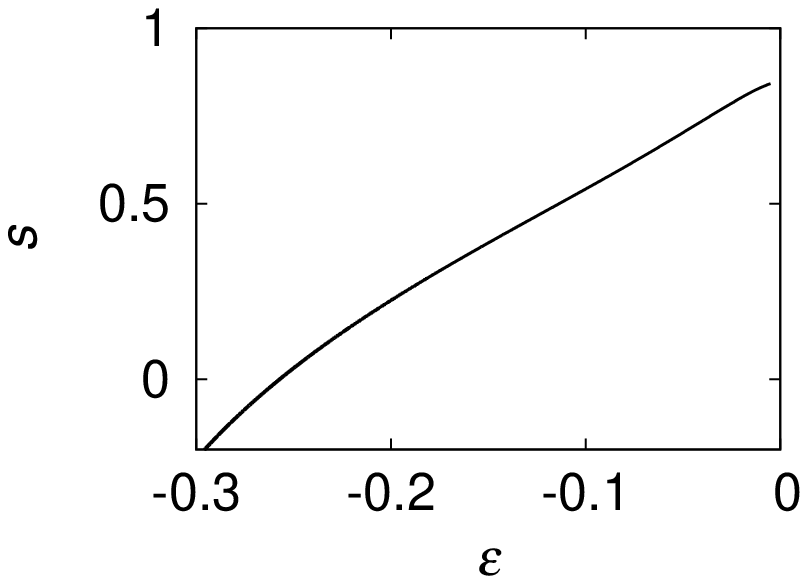}}} 
    \end{tabular}
    \caption{The inverse temperature $\beta\left(\epsilon \right)$ (Fig. 5(a)) and the entropy density $s\left(\epsilon \right)$ (Fig. 5(b)) in the microcanonical ensemble for $p=5$. $\beta\left(\epsilon \right)$ is not a monotonically decreasing function and $s\left(\epsilon \right)$ is not concave.}
    \label{figure 5}
  \end{center}
\end{figure}
On the other hand, when $p=5$, the energy density $\epsilon$ has multiple solutions for a fixed value of $\beta$ in a certain range as seen in Fig. 5(a). In this region, the entropy becomes a nonconcaved function (Fig. 5(b)) and the specific heat takes negative values, which is a clear indication of ensemble inequivalence. In contrast to previous studies$^{21-23)}$, the microcanonical entropy for $p=5$ smoothly changes from the SG solution to the PM solution at $\epsilon = -0.82$.   

\subsection{Phase diagram}
Let us next analyze the phase diagram on the $(T, \xi_{0})$ plane for $p=1$ and $p=5$.  For $p=1$, it is well known that only second-order phase transitions exist.\cite{24,25} In this case, we do not observe ensemble inequivalence. 
\begin{figure}[htbp]
    \begin{center}
      \resizebox{100mm}{!}{\includegraphics{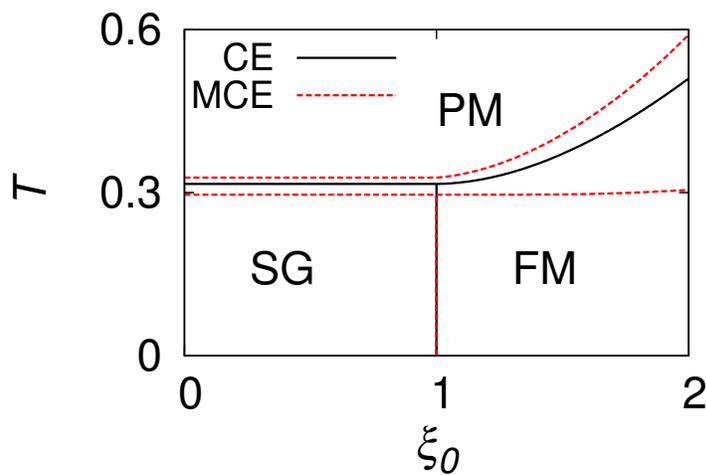}}
      \caption{The canonical (CE) and microcanonical (MCE) phase diagrams are denoted in the black solid line and the red dashed line, respectively, for $p=5$. The PM phase exists in the upper part. The SG and FM phases exist in the lower part for $\xi_{0} <1$ and $\xi_{0} >1$, respectively. In the microcanonical phase diagram, the phase is not uniquely specified for a given value of $T$ around the boundary between the PM and SG/FM phases.}
      \label{figure 6} 
    \end{center}
\end{figure}

By contrast, differences emerge for $p=5$ as shown in Fig. 6. In the canonical ensemble (the black solid lines), the phase transitions between the PM and SG or FM phases is of first order as explained $\S$ 4.1 and the phase transition between the SG and FM phases is of second order. In the microcanonical ensemble (the red dashed lines), there are regions where the phase is not determined uniquely by $T$ which is not the control parameter of the microcanonical ensemble. For $\xi_{0}<1$, as the energy density is decreased, the temperature decreases monotonically down to $T=0.29$. When the temperature reaches the value on the lower red dashed line (the line at $T=0.29$), the temperature turns to increase to $T=0.32$ (the upper dashed line) and the state changes from the PM phase to the SG phase at $T=0.31$. This non-monotonic behavior of temperature means that the specific heat takes negative values. After the temperature reaches $0.32$, the temperature turns to decrease again down to $T=0$. Such a result is observed also between the PM and FM phases for $\xi_{0}>1$. These observations mean that the state of the phase is not determined by the temperature in the range $0.29 \leq T \leq 0.32$ on the microcanonical phase diagram for $\xi_0 \le 1$, and similarly for $\xi_0>1$. This behavior has been pointed out in other cases as reported in refs. $21$, $22$ and $23$.     

\section{Conclusion}
We have derived the canonical free energy and the microcanonical entropy of the spherical spin glass model with nonlinear interactions and have shown that ensembles are not always equivalent. In particular, we have studied the behavior of the canonical free energy and the microcanonical entropy for $p=1$ and $p=5$. For $p=5$, the microcanonical entropy is not concave in a certain parameter range. As a result, the specific heat can take negative values. In addition, we analyzed the phase diagrams for $p=1$ and $p=5$ and have shown differences in the phase diagrams on the $(T, \xi_{0})$ plane for $p=5$ between the canonical and microcanonical ensembles. 

An important advantage of our study is that we have exactly solved the highly non-trivial model of spin glasses both by the canonical and the microcanonical ensembles without recourse to the replica method. We have thus unambiguously established the presence of ensemble inequivalence when quenched disorder exists in long-range exchange interactions. We expect that our present work paves the way toward further understanding of ensemble inequivalence in systems with disorder.

\section*{Appendix: Derivation of the Free energy and Entropy}
In this appendix, we derive the free energy densities and the inverse temperatures of the canonical ensemble, eqs. (13) - (19), from the saddle point equations (8) - (10) and also obtain the entropy densities and the energy densities, eqs. (25) - (30).

First, we derive eqs. (13) and (14). We evaluate the integral of the saddle point equations (8) and (10) by using the density of eigenvalues $\mu\left( \xi_{\alpha} \right)$ for $\xi_{0} <1$ and obtain
\begin{align}
1 &= \frac{1}{\lambda^{2}}\left[ z-\sqrt[]{z^{2} -\lambda^{2} \mathstrut}\right],  \tag{A.1}\\
\rho &= \frac{1}{2 \lambda^{3}}\left[ 2z^{2}-2z\ \sqrt[]{z^{2} - \lambda^{2} \mathstrut}-\lambda ^{2}\right]. \tag{A.2}
\end{align}
When $\lambda <1$, we can solve eq. (A.1) for $\lambda$ and express $\rho$ as a function of $z$,
\begin{align}
\lambda = \sqrt[]{2z-1\mathstrut},\ \rho = \frac{1}{2}\sqrt[]{2z-1\mathstrut}, \tag{A.3}
\end{align}
and obtain the inverse temperature $\beta\left( z \right)$ of eq. (14) by inserting eq. (A.3) into eq. (9). We determine the range of $z$ by the restriction $\lambda <1$ as $1/2 \leq z<1$. In this range, the summation on the right hand side of eq. (7) is estimated as follows,
\begin{align}
\frac{1}{N} \sum_{\alpha}\log\left( z-\frac{1}{2}\lambda \xi_{\alpha}\right) &\rightarrow \int^{2}_{-2}d\xi_{\alpha}\ \mu_{0} \left( \xi_{\alpha} \right) \log\left( z-\frac{1}{2}\lambda \xi_{\alpha}\right)  \nonumber \\
&= z-\log 2 -\frac{1}{2}. \tag{A.4}
\end{align} 
Thus, the free energy density and the inverse temperature of the PM phase are given by eqs. (13) and (14).

Next, we consider the saddle point equations for $\lambda \geq1$. To derive the free energy density and the inverse temperature in the SG phase, we should examine the appropriateness of integral approximation of eq. (11). As in the case of Bose-Einstein condensation, we separate the contributions of the largest eigenvalue $\xi_{\alpha} =2$ and the rest in eq. (8) as
\begin{align}
1 &= \frac{1}{N}\frac{1}{2z-\lambda \xi_{\alpha_{0}}} + \frac{1}{N}\sum_{\alpha \neq \alpha_{0}}\frac{1}{2z-\lambda \xi_{\alpha}} \nonumber \\
   &\rightarrow \frac{1}{2N}\frac{1}{z-\lambda } + \int_{-2}^{2}d\xi_{\alpha} \frac{\mu_{0} \left( \xi_{\alpha}\right)}{2z-\lambda \xi_{\alpha}}. \tag{A.5}
\end{align} 
Let us suppose that the denominator of the first term in eq. (A.5) is of order $1/N$,
\begin{align}
z = \lambda+ \frac{1}{2Nc} \simeq \lambda. \tag{A.6}
\end{align}
Then the saddle point equation, eq. (8), for $\lambda \geq1$ (i. e. $z \geq1$) is written as
\begin{align}
1 = c + \int_{-2}^{2}d\xi_{\alpha} \frac{ \mu_{0} \left( \xi_{\alpha}\right) }{ 2\lambda -\lambda \xi_{\alpha} } = c + \frac{1}{\lambda} \simeq c+\frac{1}{z}. \tag{A.7}
\end{align}
In a similar way, we evaluate eq. (10),
\begin{align}
\rho &= \frac{1}{2N}\frac{\xi_{\alpha_{0}}}{2z-\lambda \xi_{\alpha_{0}}} + \frac{1}{2N}\sum_{\alpha \neq \alpha_{0}}\frac{\xi_{\alpha}}{2z-\lambda \xi_{\alpha}}  \nonumber \\
     &\rightarrow c + \frac{1}{2}\int_{-2}^{2}d \xi_{\alpha}\ \mu_{0} \left(  \xi_{\alpha} \right) \frac{\xi_{\alpha}}{2\lambda -\lambda \xi_{\alpha}} \nonumber \\
     &= c + \frac{1}{2\lambda} \simeq 1- \frac{1}{2z}.  \tag{A.8}
\end{align}
Here we eliminate $c$ using eq. (A.7) to express $\rho$ as a function of $z$. We obtain the inverse temperature in the SG phase of eq. (16) by substituting eqs. (A.6) and (A.8) into eq. (9). To derive the free energy density in the SG phase, eq. (15), we estimate the summation term in the partition function using $z= \lambda$,
\begin{align}
\frac{1}{N} \sum_{\alpha \neq \alpha_{0}}\log\left( \lambda-\frac{1}{2}\lambda \xi_{\alpha}\right) &\rightarrow \int^{2}_{-2}d \xi_{\alpha}\ \mu_{0}\left( \xi_{\alpha} \right) \log\left( \lambda -\frac{1}{2}\lambda \xi_{\alpha}\right) \nonumber \\ 
& \simeq  \log z-\log2 +\frac{1}{2}, \tag{A.9}
\end{align}
where we notice that the contribution from the term of $\xi_{\alpha_{0}} =2$ vanishes in the thermodynamic limit. As a result, the free energy density and the inverse temperature in the SG phase are given by eqs. (15) and (16). 

Finally, we derive eqs. (17) - (19) by taking account of the ferromagnetic bias. When $\xi_{0}>1$, the isolated eigenvalue $\xi_{m}=\xi_{0}+1/\xi_{0}$ appears. The saddle point equation of eq. (8) in the FM phase is written as
\begin{align}
1 &= \int_{-\infty}^{\infty}d \xi_{\alpha}\frac{\mu \left( \xi_{\alpha}\right)}{2z-\lambda \xi_{\alpha}} \nonumber \\
   &= \frac{1}{N}\frac{1}{2z-\lambda \xi_{m}} + \int_{-2}^{2}d \xi_{\alpha}\ \frac{\mu_{0}\left( \xi_{\alpha}\right)}{2z-\lambda \xi_{\alpha}}. \tag{A.10} 
\end{align} 
The ferromagnetic ordering appears as a result of the first term in the final expression of (A.10). Assuming that
\begin{align}
2z = \lambda \xi_{m} + \frac{1}{Nd} \simeq \lambda \xi_{m}, \tag{A.11} 
\end{align} 
we can evaluate eq. (A.10) as
\begin{align}
1&= d + \int_{-2}^{2}d \xi_{\alpha} \frac{\mu_{0} \left( \xi_{\alpha} \right)}{\lambda \xi_{m}-\lambda \xi_{\alpha}} \nonumber \\
  &= d + \frac{\xi_{m}}{ 4z }\left[ \xi_{m}-\sqrt[]{\xi_{m}^{2}-4 \mathstrut} \right]. \tag{A.12} 
\end{align}
The saddle point equation of $\rho$, eq. (10), for the FM phase is
\begin{align}
\rho &= \frac{1}{2}\int_{-\infty}^{\infty}d\xi_{\alpha}\ \mu\left( \xi_{\alpha}\right) \frac{\xi_{\alpha}}{2z-\lambda \xi_{\alpha}} \nonumber\\
&= \frac{1}{2N} \frac{\xi_{m}}{2z-\lambda \xi_{m}} + \frac{1}{2}\int_{-2}^{2}d\xi_{\alpha}\ \mu_{0} \left( \xi_{\alpha}\right) \frac{\xi_{\alpha}}{2z-\lambda \xi_{\alpha}} \nonumber\\
&= \frac{\xi_{m}d}{2} + \frac{1}{2} \int_{-2}^{2}d\xi_{\alpha}\ \mu_{0} \left( \xi_{\alpha} \right) \frac{\xi_{\alpha}}{\lambda \xi_{m}-\lambda \xi_{\alpha}} \nonumber \\
&\simeq \frac{\xi_{m}}{2}\left( 1-\frac{1}{2z}\right), \tag{A.13}
\end{align}
where we have eliminated $d$ using eq. (A.12) to obtain the final expression. The inverse temperature of eq. (18) is given by plugging eqs. (A.11) and (A.13) in the saddle point equation (9). We can obtain the free energy density eq. (17) using
\begin{align}
\frac{1}{N} \sum_{\alpha}\log \left(\frac{1}{2}\lambda \xi_{m}-\frac{1}{2}\lambda \xi_{\alpha}\right) &\rightarrow \int^{2}_{-2}d\xi_{\alpha}\ \mu_{0}\left( \xi_{\alpha} \right) \log \left(\frac{1}{2}\lambda \xi_{m}-\frac{1}{2}\lambda \xi_{\alpha}\right) \nonumber \\ 
&\simeq  \log z-\log 2 +\frac{1}{2}-2h\left(\xi_{m}\right). \tag{A.14}  
\end{align}

Next, to show eq. (20), we write
\begin{align}
\Omega &=\int \frac{dt}{2\pi} {\rm tr}\ e^{it\left( E - H \right)} \delta \left( N-{\bi S}^{2} \right) \nonumber \\
            &= \frac{1}{2\pi} \int dt dz d \rho\ {\rm tr} \exp \left\{ it\left( E+N V \left( \rho \right) \right) + Nz - z\sum_{i} S_{i}^{2}  \right\} \delta \left( N\rho - \sum_{i<j} \xi_{ij}S_{i}S_{j}\right)  \nonumber \\
            &= \frac{1}{2\pi} \int dt dz d\rho d\lambda\ \exp \left\{ N \left[it\left(\epsilon +V \left( \rho \right) \right) + z - \rho \lambda - \frac{1}{N} \log  {\rm tr} \exp \left( -z\sum_{i} S^{2} +\lambda \sum_{i<j} \xi_{ij}S_{i}S_{j} \right) \right]  \right\}, \nonumber \\ \tag{A.15} 
\end{align}
and which is eq. (20) when we diagonalize random matrix $\left\{ \xi_{ij} \right\}$ and carry out the spin trace as a Gaussian integration.

The entropy densities and the energy densities in the microcanonical ensemble, eqs. (25) - (30), are derived similarly as before by the observation that $it$ plays a similar role to $\beta$ in the microcanonical case. Notice that we do not have to express $it$ as a function of $z$ in the microcanonical case, because the control variable of the system is not $it$ but $\epsilon$. Therefore, we only have to obtain $\epsilon$ as a function of $z$ using eq. (24) to derive the entropy densities and the energy densities of eqs. (25) - (30).

\end{document}